# Prediction With Dimension Reduction of Multiple Molecular Data Sources for Patient Survival

Adam Kaplan and Eric F Lock

Division of Biostatistics, School of Public Health, University of Minnesota, Minneapolis, MN, USA.

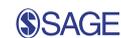

**ABSTRACT:** Predictive modeling from high-dimensional genomic data is often preceded by a dimension reduction step, such as principal component analysis (PCA). However, the application of PCA is not straightforward for *multisource* data, wherein multiple sources of 'omics data measure different but related biological components. In this article, we use recent advances in the dimension reduction of multisource data for predictive modeling. In particular, we apply exploratory results from Joint and Individual Variation Explained (JIVE), an extension of PCA for multisource data, for prediction of differing response types. We conduct illustrative simulations to illustrate the practical advantages and interpretability of our approach. As an application example, we consider predicting survival for patients with glioblastoma multiforme from 3 data sources measuring messenger RNA expression, microRNA expression, and DNA methylation. We also introduce a method to estimate JIVE scores for new samples that were not used in the initial dimension reduction and study its theoretical properties; this method is implemented in the R package R.JIVE on CRAN, in the function jive.predict.

**KEYWORDS:** Multisource data, principal component analysis, joint and individual variation explained, batch effects, survival analysis, new sample prediction, high dimensional

**RECEIVED:** April 11, 2017. **ACCEPTED:** June 8, 2017.

**PEER REVIEW:** Three peer reviewers contributed to the peer review report. Reviewers' reports totaled 604 words, excluding any confidential comments to the academic editor.

**TYPE:** Methodology

**FUNDING:** The author(s) disclosed receipt of the following financial support for the research, authorship, and/or publication of this article: This work was supported by the National Institutes of Health (grant ULI RR033183/KL2 RR0333182).

**DECLARATION OF CONFLICTING INTERESTS:** The author(s) declared no potential conflicts of interest with respect to the research, authorship, and/or publication of this article.

**CORRESPONDING AUTHOR:** Eric F Lock, Division of Biostatistics, School of Public Health, University of Minnesota, 420 Delaware Street SE, Minneapolis, MN 55455, USA. Email: elock@umn.edu

## Introduction

Dimension reduction methods are invaluable for the analysis of genomics data and other high-dimensional 'omics data. In particular, principal component analysis (PCA) and related methods reduce a large number of variables (eg, genes or genetic markers) to a small number of latent components that explain much of the variation in those variables. Principal component analysis can be used as an exploratory tool, but the principal components can also be used as predictors in a statistical model for an outcome. This latter use of PCA can solve issues of overfitting, identifiability, and collinearity that arise when a predictive model is fit using the original data. The use of principal components in linear regression and other predictive models has a long history in statistics[1] and has more recently been used in clinical cancer research to predict patient outcomes such as survival or recurrence from high-dimensional molecular predictors.[2–5] This article addresses the task of predicting an outcome from multiple sources of 'omics data, representing different but related biological components. This scenario is increasingly encountered in cancer research and other fields. For example, we consider using multisource genomic data for glioblastoma multiforme (GBM) patients from The Cancer Genome Atlas[6,7] to predict patient survival. We use 3 sources of data, capturing DNA methylation, microRNA (miRNA) expression, and gene (messenger RNA [mRNA]) expression. Long-term survival after GBM diagnosis is rare, with a median survival time of approximately 14 months with treatment.[8] Current understanding of the molecular contribution to differences in survival outcomes is limited but suggests that miRNA, DNA methylation, and the regulation of gene expression all play a key role.[9,10]

The application of classical dimension reduction techniques such as PCA for prediction from multisource genomic data is not straightforward. A sensible ad hoc approach is to perform a separate dimension reduction for each source and then combine the source-specific components in a predictive model; this approach is explored in the work by Zhao et al[11] to predict survival from mRNA, miRNA, methylation, and copy number aberration data. However, components extracted from related sources may be collinear or redundant. At the other extreme, one could concatenate the data sources and use PCA or other dimension reduction approaches on the combined data (see, eg, consensus PCA[12]) prior to predictive modeling. However, this approach lacks interpretability regarding the contribution of each data source and may not capture relevant signals that are specific to a single data source precisely.

There is a nascent literature in several computational domains on the integrative dimension reduction of multisource data.[13–16] These methods identify lower-dimensional structure that is shared across multiple sources and structure that is specific to each source. In particular, Joint and Individual Variation Explained (JIVE)[17,18] is a direct extension of PCA for multisource data. These methods have been used for exploratory analysis but not for predictive modeling. A recent survey of integrative analysis methods for cancer data identified many exploratory methods for multisource integration, and many predictive methods for a single source, but no methods for prediction from multisource data.[19] In this article, we describe an approach in which JIVE components are used for predictive modeling, evaluating the properties and predictive accuracy of JIVE in a statistical modeling framework. Joint and Individual Variation Explained identifies a small number of joint latent components across all data sources, and individual components specific to each data source, that maximize the overall variation explained. Thus, redundancy and collinearity among the sources are accounted for in the joint components, and the individual components facilitate interpretation of the unique contribution of each data source in a predictive model.

The purpose of this article is 2-fold:

1. To illustrate the advantages of JIVE and multisource dimension reduction more broadly for predictive modeling of high-dimensional multisource data.
2. To introduce a method for the prediction of JIVE component scores for new samples that were not included in the original JIVE dimension reduction step, called jive. predict.

We also describe data preprocessing and postprocessing steps to improve predictive performance.

The rest of this article is organized as follows. In section "Methods," we review the JIVE algorithm and explain its use in statistical modeling. In section "Simulations," we discuss 2 simulations to illustrate and assess the efficacy of using JIVE in a predictive modeling framework. In section "GBM Data Application," we describe an application to predict patient survival for the multisource GBM data, using JIVE components in a Cox proportional hazards model; we also discuss preprocessing and postprocessing steps and compare with separate and concatenated PCA approaches. In section "Post hoc prediction," we introduce jive.predict and explore its theoretical properties and application to the GBM data. Section "Discussion" concludes with a discussion and summary.

## Methods

### Principal component analysis

Here, we briefly describe the mathematical form and notation for PCA, before describing JIVE as an extension to multisource data in section "Joint and Individual Variation Explained." Let $X : p \times n$ be a data matrix, with $p$ variables (rows) measured for $n$ samples (columns). The variables are centered to have mean $0$, to remove baseline effects. The first $r$ principal components give a rank $r$ approximation of $X$:

$$X \approx US$$

where $U : p \times r$ are the variable *loadings* indicating the contribution of the variables to each component, and $S : r \times n$ are the sample *scores* that are used as an $r$-dimensional representation for the samples. The scores and loadings are chosen to minimize the sum of squared residuals $||X - US||_F^2$, where $||\cdot||_F$ defines the Frobenius norm. Thus, $U$ and $S$ can also be obtained from a rank $r$ singular value decomposition of $X$; the columns of $U$ are standardized to have unit norm, and both the columns of $U$ and the rows of $S$ are orthogonal.

Concatenated PCA describes the application of PCA to multiple sources of data:

$$X_1 : p_1 \times n, X_2 : p_2 \times n, \ldots, X_m : p_m \times n. \quad (1)$$

Each data source is standardized separately to have the same total sum of squares, to resolve scale differences between the sources. The sources are then concatenated together by the rows to form $\mathbf{X} : [X_1^T \ldots X_m^T]$, and PCA is performed on the concatenated data $\mathbf{X}$. This method estimates the joint variability structure for multiple sources but does not allow for estimation of the individual sources' structure. Individual PCA corresponds to a separate PCA factorization for each source.

### Joint and Individual Variation Explained

Joint and Individual Variation Explained was originally developed as a data exploration technique for measuring patterns shared among and expressed within multiple data sources. Joint and Individual Variation Explained is an extension of PCA, to allow for more than 1 high-dimensional data source. Specifically, JIVE serves as a compromise between concatenated PCA and individual PCA. For multisource data (equation (1)), the JIVE factorization is as follows:

$$X_1 = \underbrace{U_1 S}_{J_1} + \underbrace{W_1 S_1}_{A_1} + \text{Error}_1$$
$$\vdots$$
$$X_m = \underbrace{U_m S}_{J_m} + \underbrace{W_m S_m}_{A_m} + \text{Error}_m$$

where $S : r \times n$ are the joint sample scores and $U_i : p_i \times r$ are loadings for the joint scores in the $i$th data source. Thus, the form of this joint approximation $\mathbf{J} = [J_1^T \ldots J_m^T]$ is equivalent to a concatenated PCA with $r$ components. The $S_i : r_i \times n$ are sample scores capturing variation specific source $i$, and $W_i : p_i \times n$ are the corresponding loadings; thus, the form of the individual structure $\mathbf{A} = [A_1^T \ldots A_m^T]$ is equivalent to individual PCA for each source. Orthogonality between the rows of $J_i$ and $A_i$ is necessary to uniquely define the joint and individual structure. In addition, one can optionally enforce the constraint that the $A_i$'s are orthogonal to each other. The ranks of the model (ie, number of joint and individual components) are estimated via a permutation testing approach. For given ranks, the estimates are obtained via an iterative algorithm to minimize the overall Frobenius norm of the error:

$$\sum_{i=1}^{m} ||\text{Error}_i||_F^2$$

For more computational and theoretical details on the JIVE algorithm, and its use in an exploratory context, we refer to Lock et al[17] and O'Connell and Lock.[18]

## JIVE for prediction

Consider predicting an outcome variable $Y_k$ for $k = 1,\ldots,n$, from multisource data (equation (1)). The entire data $\mathbf{X}$ has dimension $p = \sum_{i=1}^{m} p_i$ for each sample, and each $p_i$ is commonly in the thousands for genomic data. Through JIVE, these data are transformed to a small number of jointly present or source-specific latent patterns $(r + \sum_{i=1}^{m} r_i) \ll p$. We use the JIVE scores, $S$ and $S_i$, for further statistical modeling. This alleviates issues of overfitting and multicollinearity, as the rows of $S$ and each $S_i$ are orthogonal. Straightforward linear model using all joint and individual components is as follows:

$$Y_k = \beta_0 + \sum_{v=1}^{r} \beta_v S_{vk} + \sum_{i=1}^{m} \sum_{j=1}^{r_i} \beta_{ij} S_{ijk}$$

The joint component scores $S_{vk}$ account for collinearity and redundancy among the different sources, whereas the individual scores $S_{ivk}$ give the unique contributions from each data source.

For further interpretation, it helps to consider the relative contribution of each variable (eg, gene or miRNA) to the fitted model, and these can be obtained via the loadings for each component. For this purpose, we combine the loadings for the joint and individual components for a given data source, $U_i$ and $W_i$, and weigh them according to the estimated model coefficients $\hat{\beta}$ to obtain *meta-loading* $\phi_i : p_i \times 1$:

$$\phi_i = \sum_{v=1}^{r} \beta_v \mathbf{u}_{iv} + \sum_{j=1}^{r_i} \beta_{ij} \mathbf{w}_{ij}$$

where $\mathbf{u}_{iv}$ is the $v$th column of $U_i$, and $\mathbf{w}_{ij}$ is the $j$th column of $W$. These meta-loadings approximate the relative overall size and direction of the contribution for each variable and may be used to evaluate candidate variables or data sources.

## Simulations

In this section, we use JIVE scores for prediction in 2 illustrative simulations: one for modeling a binary outcome and another for a time-till event outcome. The strengths of JIVE are highlighted in this section by illustration: estimating shared data structures across data sources and the individual sources of signal that each data source independently provides.

### Binary outcome

In this section, we consider a simulation design with a binary group structure analogous to that in Section 4.10.3 in the study by Lock.[20] We will use this simulation to illustrate the use of JIVE scores to distinguish sample groups in an exploratory fashion and their use in a predictive model for group membership. First, we created 2 data sources, $X_1$ and $X_2$, each with dimension $5000 \times 300$ and independent $Normal(0, 2)$ entries as background noise. The shared columns (samples) belong to 2 classes: 150 in Class +1 and 150 in Class −1. For joint signal, we added 1 to a set of 50 samples in Class +1 and subtracted 1 from a set of 50 samples across 50 rows in both $X_1$ and $X_2$; thus, these samples are distinguished on both data sources. For individual signal in $X_1$, we added 1 to another set of 50 columns in Class +1 and subtract 1 from another set of 50 columns in Class −1, for 50 rows in $X_1$ only. For individual signal in $X_2$, we added 1 to the remaining 50 columns in Class +1 and subtract 1 from the remaining 50 columns in Class −1, for 50 rows in $X_2$ only. This simulation design is visually represented in Figure 1.

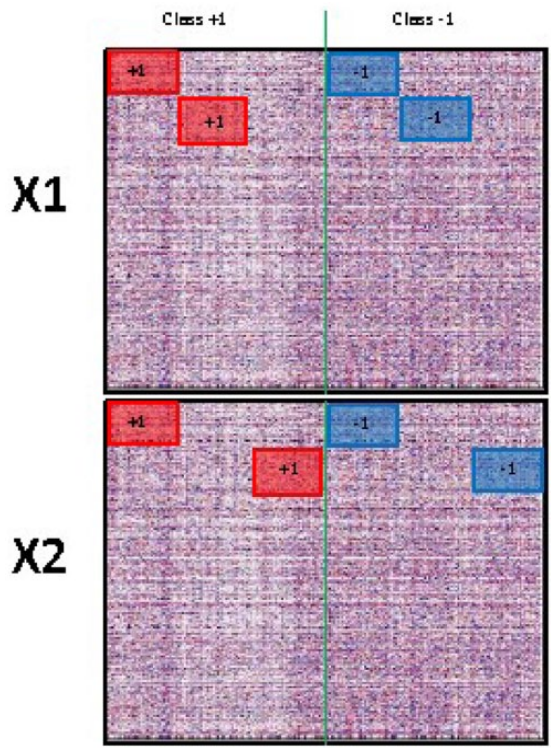

**Figure 1.** Simulation design for binary outcome. Reproduced with permission from Lock.[20]

For a concrete example of how JIVE can capture the information even when reducing to ranks of 1, we consider the plot of the joint and individual scores identified by JIVE in Figure 2. This figure shows scatterplots of the scores for joint, individual 1, and individual 2, against each other. For example, the plot of joint against individual 1 in the top left panel accurately separates the corresponding sources of signal along the axes. The joint source is split along the vertical, and the individual 1 source is picked up and split against the $x$-axis; the samples that have signal only in source 2 are not well distinguished.

We next use the joint, individual 1, and individual 2 scores as covariates to predict class membership, using a logistic model:

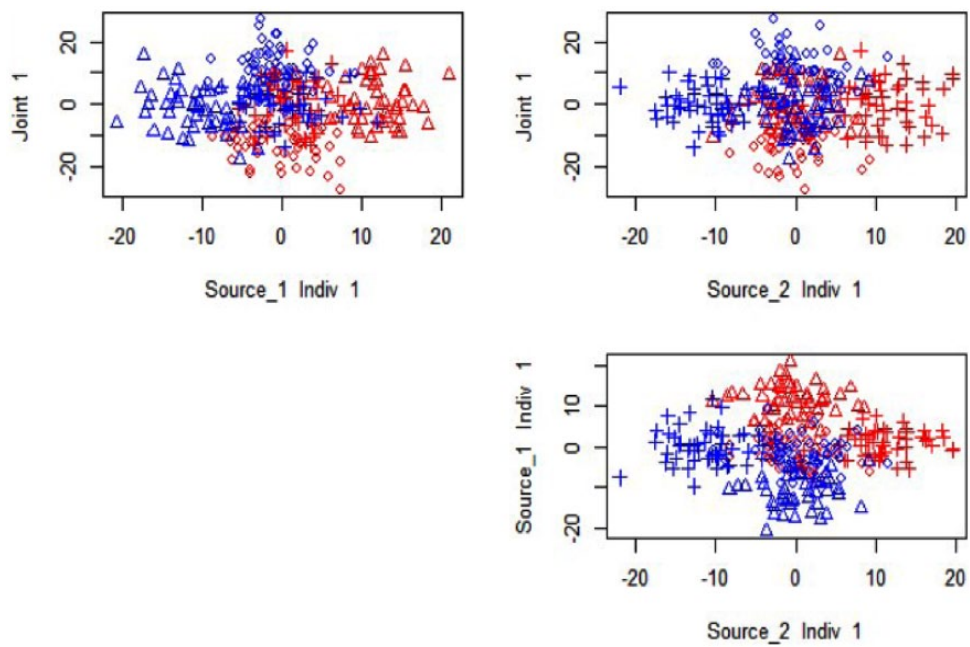

**Figure 2.** Joint and Individual Variation Explained results for 1 joint rank and 1 individual rank for each source: ◊ indicates jointly distinguished, △ is distinguished in source 1 only, and + is distinguished in source 2 only; blue indicates Class −1, whereas red indicates Class +1.

$$\text{Logit}(\pi) = \alpha + \beta_1 \mathbf{J}_i + \beta_2 \mathbf{I.1}_i + \beta_3 \mathbf{I.2}_i$$

where $\mathbf{J}$, $\mathbf{I.1}$, and $\mathbf{I.2}$, are the joint, individual 1, and individual 2 source scores, respectively, and $\pi$ is the probability that the sample is from Class +1. After running a simple logistic model, the coefficients estimated are given in Table 1.

All the estimates were highly significant for discriminating Class +1 and Class −1 ($P<.0001$) and had similar $z$ scores representing the relative effect size. This demonstrates that shared signal in $X_1$ and $X_2$ and complementary information in $X_1$ and $X_2$ all play an important role in the distinction of the 2 classes. The sign of each coefficient is not directly interpretable without further investigation because the scores and loadings are identifiable only up to a sign transformation, eg, $\tilde{W}_i \tilde{S}_i = W_i S_i$ if $\tilde{W}_i = -W_i$ and $\tilde{S}_i = -S_i$.

Fitted value and residuals plots are shown in Figure 3 and are for the most part as expected. The Class +1 signal is associated with higher predicted probabilities, and the class −1 signal is associated with lower predicted probabilities. The probability of being in the positive signal class tends to decrease with an increase in joint scores, and the probability of being in the positive signal class tends to increase with each increase in individual 1 and individual 2 scores.

We assess the accuracy of the JIVE prediction approach with 100 randomly generated data sets under the above simulation scenario. Within each simulation, we ran an N-fold (leave-one-out) cross-validation and obtained the predicted probability of the held-out sample being associated with the Class +1. Then, we computed the mean absolute error (MAE) of these predictions for each simulation. To compare predictive accuracy, we compared the results for JIVE with the correct ranks given (joint rank = 1, individual ranks = (1,1)) to the resulting model fit using the components of a concatenated (joint) or separate (individual) PCA analyses. To capture all of the structure in the data, rank 3 was used for concatenated PCA (joint rank = 3, individual ranks = (0,0)), and rank 2 was used for individual PCA (joint rank = 0, individual ranks = (2,2)). The simulation mean and standard error of the MAE for JIVE, concatenated PCA, and individual PCAs were 0.1576 (0.00299), 0.2523 (0.00511), and 0.2165 (0.00339), respectively. Joint and Individual Variation Explained attained the lowest error in probability prediction over the simulations while also obtaining the smallest variability for the errors.

**Table 1.** Logistic regression results using Joint and Individual Variation Explained scores.

|  | JOINT | INDIVIDUAL 1 | INDIVIDUAL 2 |
| --- | --- | --- | --- |
| Estimate | −0.261 | 0.434 | 0.395 |
| Standard error | 0.041 | 0.061 | 0.061 |
| $z$ score | −6.378 | 7.084 | 6.516 |
| $P$ value | <.0001 | <.0001 | <.0001 |

### Time-till-event outcome

In this example, we consider using JIVE for predictive modeling of time-till-event data. Using the same data-generating scheme for 2 sources $X_1$ and $X_2$ described in section "Binary outcome," we assume that the 2 classes {−1, 1} correspond to patients with different survival patterns. The samples of Class −1 are assigned a random time from the Gamma(1, 1) distribution. Similarly, we assigned a random survival time from the Gamma(3, 1) distribution to the samples in Class +1. Then, we censor all event times greater than 3. We use a Cox proportional hazards model[21] using the JIVE scores as predictors:

$$\lambda(t \mid X_i) = \lambda_0(t) \exp(\beta_1 \mathbf{J}_i + \beta_2 \mathbf{I.1}_i + \beta_3 \mathbf{I.2}_i)$$

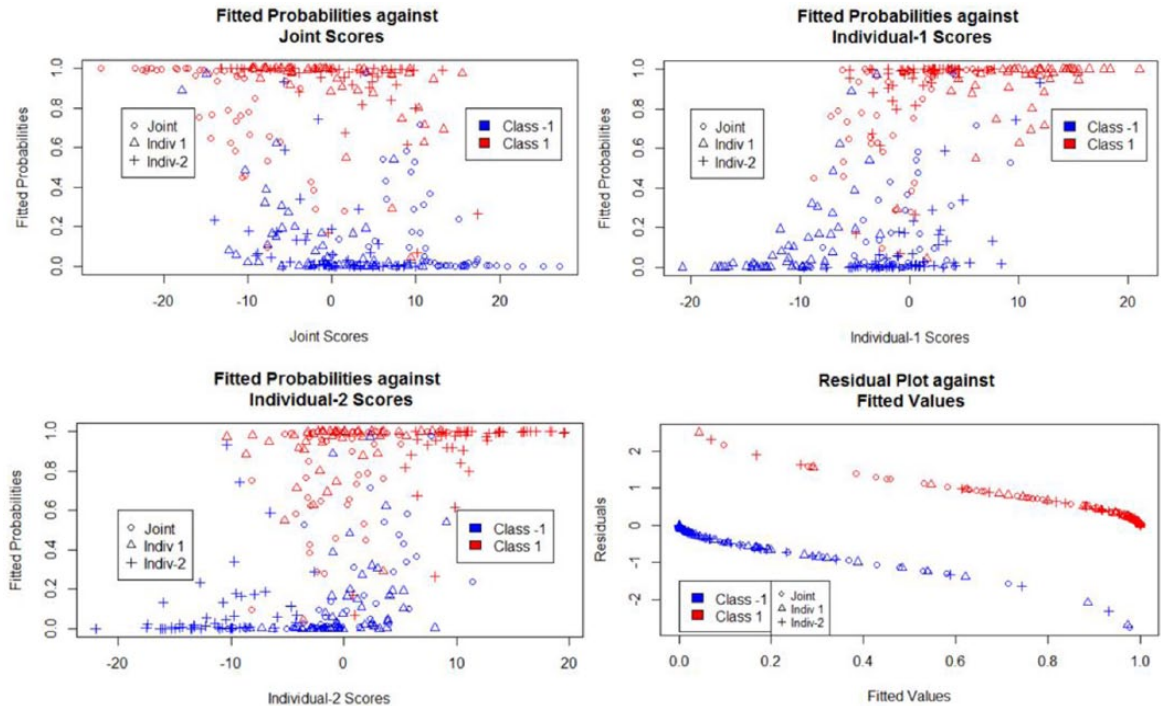

**Figure 3.** Diagnostics for logistic regression simulation.

This expression gives the hazard rate $\lambda$ at time $t$ for subject $i = 1, \ldots, 300$.

For this example, we observe strong and significant relationships between the hazard rate and all 3 components: joint, individual 1, and individual 2. To assess sensitivity of the results, we generated 100 data sets independently under the simulation scheme detailed above, computed the JIVE components for each data set, and fitted the resulting Cox model for each data set. The results are summarized in Table 2. We calculated the mean absolute coefficients, the mean absolute z score, and the standard deviation of the absolute coefficients and then created a distribution of the *P* values that were significant at the .01 level.

The results were generally robust to the random seed that generated the values for a given simulation. The standardized coefficients were generally similar across simulations, and all 3 components were significant in at least 90% of cases. In a positive or negative way, depending on the random number seed, the joint, individual 1, and individual 2 scores each yield critical information for predicting survival. We used the *survival* package in R for the Cox proportional hazards model estimation.[21]

Similar to the aforementioned binary outcome simulation, we assessed the prediction accuracy with 100

**Table 2.** Results from 100 simulations Cox proportional hazards regression results using Joint and Individual Variation Explained scores.

|  | JOINT | INDIVIDUAL 1 | INDIVIDUAL 2 |
| --- | --- | --- | --- |
| Mean \|*Coef*\| | 0.0297 | 0.0488 | 0.0483 |
| Standard error of coefficient | 0.001 | 0.001 | 0.001 |
| Mean \|*z score*\| | 4.295 | 5.346 | 5.358 |
| Standard error of *z* | 0.1376 | 0.101 | 0.107 |
| Proportion of *P*-value < 0.01 | .92 | .98 | 1.00 |

simulations of the time-till-event outcome scenario. Within each simulation, we ran an N-fold cross-validation and obtained the correlation between the predicted median survival time and the observed survival time. The simulation mean and standard error of the correlation for JIVE, concatenated PCA, and individual PCAs were 0.3395 (0.00651), 0.2797 (0.00689), and 0.3073 (0.00655), respectively. The use of JIVE with the correct ranks for prediction gives optimal performance; however, the 2 methods with misallocated ranks still give reasonable predictive accuracy.

Altogether, the binary and time-till-event simulations elucidate more into JIVE's improvement on PCA, for both interpretation and predictive accuracy. Rank misspecification (as in the concatenated and individual PCAs for both of the aforementioned scenarios) decreases prediction accuracy but can still give reasonable estimates.

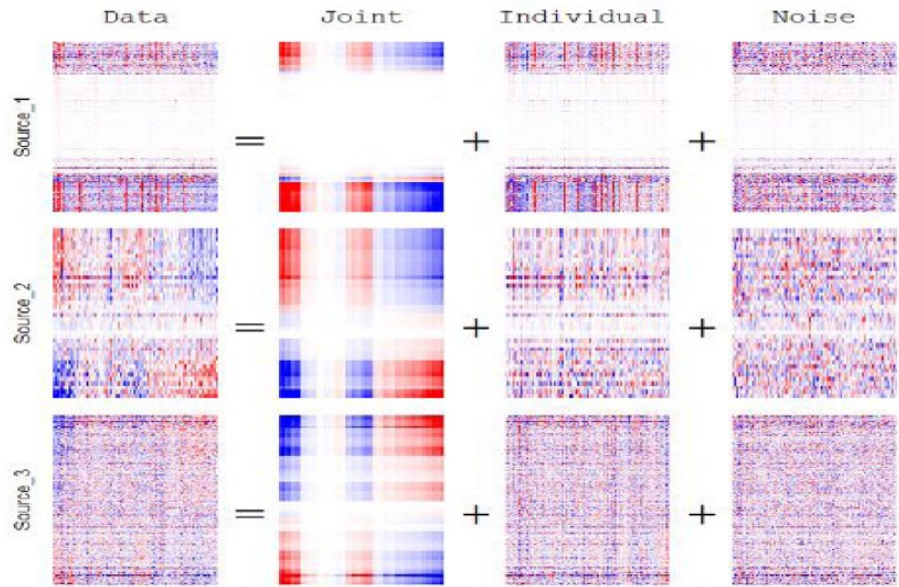

**Figure 4.** Preprocessed Joint and Individual Variation Explained heatmap for methylation, microRNA, and gene expression data, respectively, ranks: joint = 1, individual = (13, 9, 36), red is positive signal and blue is negative signal.

## GBM Data Application

In this section, we analyze the relationships of gene expression, miRNA, and DNA methylation data in relation to patient survival. We note that the DNA methylation of genes—specifically MGMT—are observed by around 50% of the patients with GBM. The methylation status of MGMT is related to the survival of those patients who underwent radiation therapy. Across many studies of GBM, patient age has a reoccurring relationship with patient survival.[9] The use of JIVE on other sources of data may reveal stronger predictions than predictions that primarily use age, and this is the article's aim.

### Preprocessing

Here, we describe preprocessing steps for the 3 data sources. Gene expression data were measured using the Agilent G450 microarrray, miRNA was measured using the Agilent $8 \times 15K$ Human miRNA array, and methylation was measured using the Illumina HumanMethylation27 BeadChip and HumanMethylation450 BeadChip platforms. We assessed batch effects and other technical artifacts via an initial PCA of each data source. Gene expression and DNA methylation had plate number effects and platform effects, respectively, that were corrected for using *ComBat*[22] using a nonparametric empirical Bayesian approach.

Following preprocessing and batch adjustment, the full data had $p_1 = 21180$ methylation sites, $p_2 = 534$ miRNAs, and $p_3 = 17814$ genes. We conducted an initial filtering step for the variables based on their univariate associations with survival, to remove irrelevant variables. This approach is similar to that described in a supervised PCA method for a single data source.[2] Specifically, we ran a univariate Cox survival model for each variable in each data source against the survival outcome. We collected the *P* values and used a false discovery rate (FDR) of 20% to collect a reduced number of variables marginally associated with the outcome.[23] The resulting data set count for the 20% FDR preprocessing step was as follows: 2700 variables for DNA methylation, 40 variables for miRNA, and 2085 variables for gene expression, for $n = 304$ common subjects.

### JIVE results

Following the preprocessing of the 3 data sources, we applied JIVE using R Studio and the package r.jive.[18] Joint and Individual Variation Explained estimated the joint rank and (3) individual ranks to be (1, 13, 9, 36), respectively. One may view these ranks as individual columns of *composite biomarkers* for each patient. Thus, a total of 59 components were considered for predictive modeling, capturing the variation patterns either within or across the 3 data sources. This significant reduction in variables accomplished by filtering and JIVE allowed for standard statistical analysis methods for 304 shared patients among the 3 data sources.

Figure 4 shows a heatmap of the estimated joint and individual structure over the 3 data sources. The white space in Figure 4 revealed that a large subset of variables (rows) in the methylation data source (Source_1) yielded no contribution to the variation within that data source. Also, from the section of the heatmap that corresponds to the estimated joint structure, there was a small subset of subjects (columns) that contributed little to no variation in the estimated joint structure. In Figure 5, the estimated scores are plotted against each other, whereas color coding according to clinical data was implemented to aid explanation in these estimated joint and individual structures. Joint scores elicited a relationship to clinical subtype. Proneural and G-CIMP, to classical and neural, finally mesenchymal, are distinguished in an increasing joint component fashion.

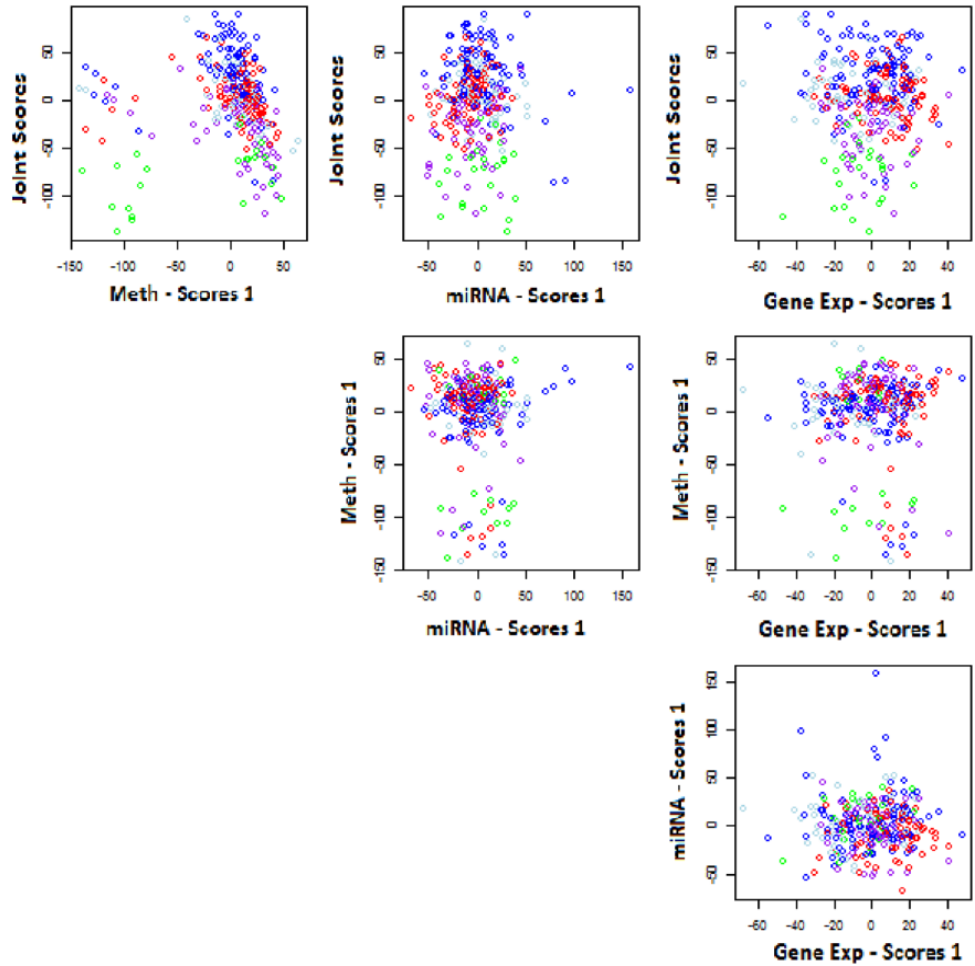

**Figure 5.** Joint and Individual Variation Explained scores colored by clinical subtype: blank, classical, G-CIMP, mesenchymal, neural, proneural—for the processed data. MiRNA indicates microRNA.

*Cox model and validation*

In addition to the JIVE scores, we also included age and sex into the Cox model as potential predictors. Thus, the full model with all variables included was as follows:

$$\lambda(t \mid X_i) = \lambda_0(t) \exp \left( \beta_1 \text{Age}_i + \beta_2 \text{Gender}_i + \underbrace{\beta_3 J_{3i}}_{\text{Joint Score}} + \underbrace{\sum_j \beta_j I.1_{ji}}_{\text{Methylation Scores}} + \underbrace{\sum_k \beta_k I.2_{ki}}_{\text{miRNA Scores}} + \underbrace{\sum_r \beta_r I.3_{ri}}_{\text{Gene Exp. Scores}} \right)$$

where $j = 1,\ldots,13$; $k = 1,\ldots,9$; $r = 1,\ldots,36$; $n = 1,\ldots,304$.

We applied a backward/forward selection process[24] based on Akaike information criterion (AIC)[25] to select the most predictive subset of JIVE components and other variables in the final model. Then, we ran the model through a series of permutation tests to assess whether the JIVE results add explanation of the variability in the survival times at all, as well as whether they are informative *in addition to* that variability already explained by the patient age. Specifically, we created a null distribution of AIC values where the relationships of the survival outcomes and all other variables are severed via permutation and compared this distribution with the AIC obtained for the true data. Second, we permuted the survival outcomes with age and the other variables in the model separately. This second strategy reflected a null distribution where variables added no information on top of age, in relation to the survival outcome. We compared the observed AIC with the null distribution of AICs created with the age permutation. Finally, to assess the estimation accuracy of the method to predict survival, we conducted N-fold (leave-one-out) cross-validation. Each fold contained a backward-forward step variable selection that found the best fitted model given the set of training samples and predicted the median survival time for the left out test sample. Method accuracy was measured using the correlation of the $\log_{10}(\cdot)$ of the prediction and observed times.

*Comparison of methods*

We used the 3 objective resampling assessment strategies described above—permutation testing, permutation testing in addition to age, and cross-validated prediction

**Table 3.** Comparing modeling results across dimension-reducing strategies.

| METHOD | AIC VALUE | PERM. AGE, P VALUE | CORR. (LOG(.)) | 95% CI |
| --- | --- | --- | --- | --- |
| JIVE | 1887.394 | <.0001 | 0.637 | 0.55–0.71 |
| Concatenated PCA | 1869.617 | <.0001 | 0.507 | 0.402–0.599 |
| Individual PCA | 1881.127 | <.0001 | 0.289 | 0.163–0.407 |
| Methylated PCA | 1920.857 | <.0001 | 0.468 | 0.358–0.565 |
| MiRNA PCA | 1955.293 | .004 | 0.364 | 0.244–0.474 |
| Gene expression PCA | 1938.709 | .004 | 0.426 | 0.311–0.529 |

Abbreviations: AIC, Akaike information criterion; PCA, principal component analysis.

accuracy—on the concatenated PCA, altogether individual PCA, and the 3 individual PCA results to compare with the predictive model obtained through JIVE. The number of components for the PCA analyses was chosen to be consistent with the estimated joint rank and individual ranks from JIVE. That is, the number of components for concatenated PCA was given by summing all joint and individual ranks. The estimated ranks for the individual PCAs were $r_{joint} + r_i$ where $r_i$ is the rank for data source $i$, estimated by JIVE. Therefore, the corresponding ranks for individual PCAs of methylation, miRNA, and gene expression were (14, 10, 37), respectively. To compare methods with JIVE, we conducted the permutation tests and the N-fold cross-validation for each of the aforementioned methods.

Table 3 displays the series of tests for each of the methods. All 3 integrative methods generally had better performance than any of the 3 models that only consider each data source separately, suggesting the need for a multisource predictive model. And, of the 3 integrative approaches, JIVE had a stronger performance under cross-validation than the other 2 ad hoc PCA approaches, with the highest correlation between predicted median survival time and the observed survival time. The AIC values for the final models selected under each strategy were similar.

The permutation $P$ values that did not permute age along with the outcomes were all $< .0001$; therefore, all of these proposed methods resulted in strong associations with GBM survival—these results are not presented in Table 3. The information provided by JIVE on top of that provided by patient age was significant, and the distribution of permutation statistics is shown in Figure 6. The observed vs predicted median survival values under cross-validation using JIVE are shown in Figure 7.

## Post Hoc Prediction

In this section, we consider prediction from a new set of multisource patient data, $\mathbf{X}^*$, that was not a part of the initial dimension reduction, given that we have already fitted a predictive model to JIVE results. In PCA, this task is straightforward, by simply projecting the new data into the previously obtained principal component loadings to obtain scores for the new data; however, for JIVE and other multisource factorizations, this is not straightforward. Thus, we develop a method called jive.predict for out-of-sample prediction from multisource data. We assume that JIVE has already been applied to a data set $\mathbf{X}$ with multisource structure (equation (1)). Joint and Individual Variation Explained gives joint and individual loadings $U$ and $W_i$ for the present data sources. Let $\mathbf{X}^*$ be measurements for the same data sources but for a new set of patients. We use the old loadings from the previous JIVE analysis to compute sample scores for the new patients, $S^*$ and $S_i^*$, that minimize the squared residuals for those patients.

### JIVE.predict—the algorithm

This problem frames the JIVE algorithm in a new light: given a new data $\mathbf{X}^*$ how can we estimate the new scores $S^*$ and $S_i^*$, where $i$ is the index of the data type. For simplicity, we will focus on the scenario of 2 data sources $\mathbf{X} = [X_1^T X_2^T]^T$, with the following estimated JIVE decomposition:

$$X_1 \approx U_1 S + W_1 S_1$$
$$X_2 \approx U_2 S + W_2 S_2$$

where $S : r \times n$, $U_i : p_i \times r$, $W_i : p_i \times r_i$, and $S_i : r_i \times n$ for $i = 1, 2$. Consider a new data set $\mathbf{X}^* = [X_1^{*T} X_2^{*T}]^T$, with the same number of variables and data sources (rows) but a different set of $n^*$ samples (columns). To approximate the scores for the new samples for predictive modeling, we use the same loadings as those for the previous JIVE decomposition:

$$X_1^* \approx U_1 S^* + W_1 S_1^*$$
$$X_2^* \approx U_2 S^* + W_2 S_2^*$$

where $S^* : r \times n^*$, $U_i : p_i \times r$, $W_i : p_i \times r_i$, and $S_i^* : r_i \times n^*$ for $i = 1, 2$. Thus, we are simply computing scores for the new samples that are consistent with the previous JIVE analysis. To do this, we iteratively solve for the scores that minimize the sum of squared error, with the previous loadings fixed. The procedure for the estimation is brief and for our examples takes about 5 to 10 iterations to converge.

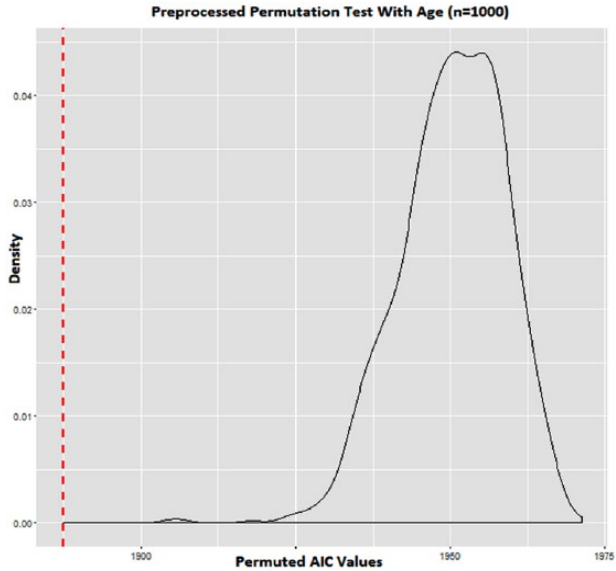

**Figure 6.** Age-outcome permutation results using Joint and Individual Variation Explained, *P* < .0001, red line is observed as AIC 1887.394.

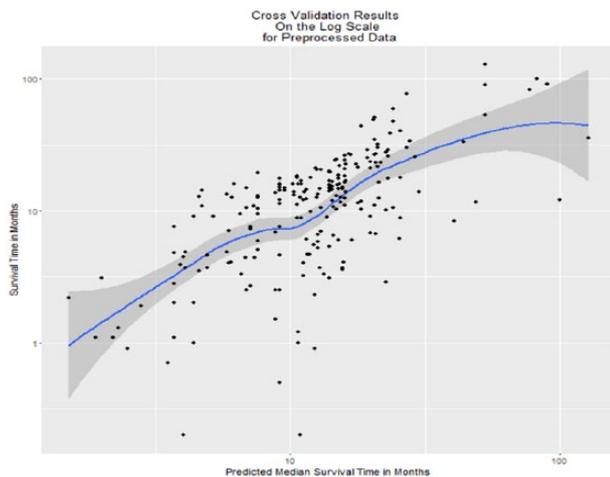

**Figure 7.** N-fold cross-validation results on the $\log_{10}(\cdot)$ scale: correlation = 0.637, *P* < .0001, 95% confidence interval: 0.55-0.71.

The algorithm is as follows:

- Initiate $S_i^* = 0$ (individual scores).

We then minimize the error by iteratively estimating $\mathbf{S}^*$ and $S_i^*$ until convergence:

$$\text{Error} = \sum_i \| X_i^* - U_i S^* - W_i S_i^* \|_F^2$$

- $\mathbf{S}^* = [U_1 U_2]^T (\mathbf{X}^* - [W_1 S_1^* W_2 S_2^*]^T)$
- $S_i^* = W_i^T (X_i^* - U_i S^*)$
- Stop until convergence.

Using this approach, the original JIVE decomposition and the scores for the new samples are both estimated via minimizing the sum of squared residuals. In fact, if the optional orthogonality between individual structures is not enforced in the original JIVE decomposition, the jive.predict scores are exactly the same as the JIVE scores for the same data. That is, if $\mathbf{X}^* = \mathbf{X}$, then $S^* = S$ and $S_i^* = S_i$ for all $i$. The formal proof of this result is given in Appendix 1. No analogous result holds if orthogonality is enforced between the individual structures, namely, $S_1^T S_2 = 0$ because the individual scores estimated via jive.predict may not be orthogonal. However, our empirical results suggest that the results of jive.predict are still reasonable even when individual orthogonality is enforced in the original JIVE decomposition.

### GBM data illustration

In this section, we illustrate jive.predict using the GBM survival data from section "GBM data application."

We replicated the scenario of the new patient prediction with a 5-fold cross-validation. For each fold, 20% of the patients were placed in the test data set, whereas 80% of the remaining patients were used as the training data set. Each patient was used in the test data once. The training data set underwent a JIVE decomposition and then used as covariates in a Cox proportional hazards model with age and sex. The decomposition was followed by a backward/forward variable selection process. The test data were not included in the original JIVE analysis or modeling, but their sample scores were predicted using jive.predict and the JIVE loadings from the training data set. To evaluate whether orthogonality between the individual scores increases model accuracy, there is another case where we also applied jive.predict on the training data using the JIVE results of that training data. Backward/forward variable selection was implemented, after the fit of the Cox proportional hazards model using the scores as covariates.

Predicted vs true median survival plots with individual orthogonality enforced are shown in Figure 8, and with individual orthogonality not enforced are shown in Figure 9. The color of each of the plotted coordinates corresponds to the fold of the cross-validation, showing that the observation in that fold had an event (death), and the predicted median survival time was not missing.

The correlations were not that different in values between the 2 cases of orthogonality. In the case of orthogonality, the correlation with orthogonality enforced between individual scores was 0.3966 (95% CI: 0.277-0.504, *P* < .0001). When orthogonality was not enforced (using *jive.predict* on the training data) the correlation was 0.3992 (95% CI: 0.28-0.506, *P* < .0001). This high value in correlation explains that the model's predictions of higher median survival times moderately correspond to higher survival times from the actual data.

An important consideration when using jive.predict is consistency in the centering and scaling approaches used for the test data $\mathbf{X}^*$ and the training data $\mathbf{X}$. We centered and scaled the scores from the JIVE results before using them as covariates. This was for ease of interpretation. The results from an application of *jive.predict* on new data $\mathbf{X}^*$ may differ depending on if $\mathbf{X}^*$ follows similar scaling and centering as the scores from the JIVE decomposition on data $\mathbf{X}$. For the 2 cases above, we scaled and centered the scores using the centering and standard deviations of the JIVE scores from the training data set. If the new data were measured or processed differently than the training data, or were different in any other systematic way, then the model predictions may be biased. Thus, when using jive.predict, it is important to ensure that the new data are collected in a way that is consistent with the old data.

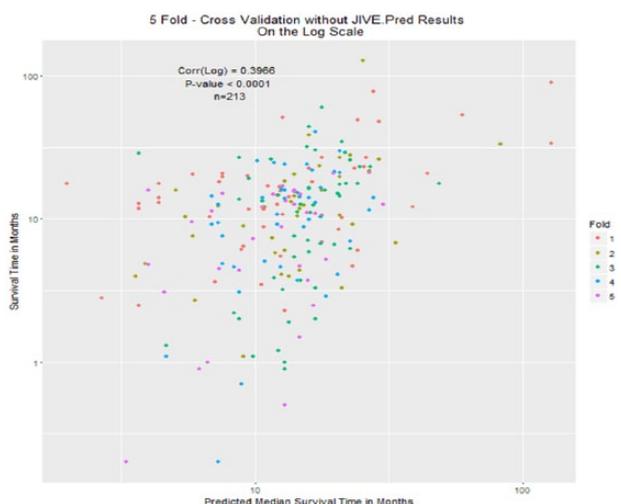

**Figure 8.** Fivefold cross-validation: test scores estimated using jive.predict, training model fit using Joint and Individual Variation Explained only.

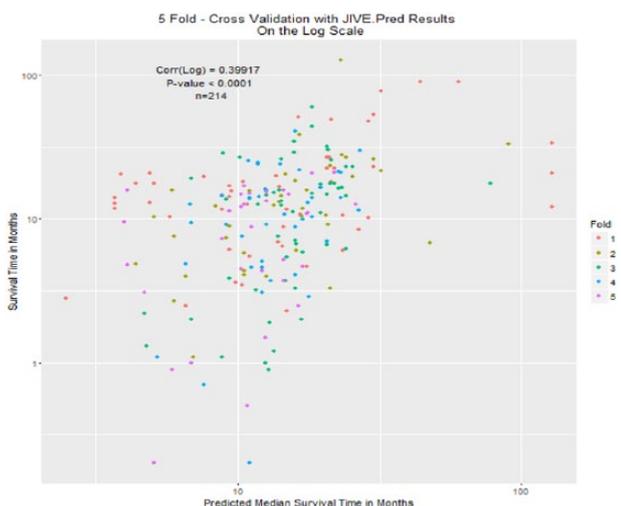

**Figure 9.** Fivefold cross-validation: test scores estimated using jive.predict, training model fit using jive.predict.

## Discussion

Joint and Individual Variation Explained addresses 2 problems: handling multiple high-dimensional data sources as well as dimension reduction assuming matrix rank sparsity. In this article, we have explored the use of such a method for statistical modeling and prediction. For GBM data sources and other data sources relating to cancers with more complex genetic factors, JIVE is a strong exploratory tool, as well as a strong predictive analysis tool for these problematic high-throughput data scenarios.

One key feature of JIVE is that this method ensures that the estimated joint and individual structures are orthogonal to each other, meaning that there is no overlap between the estimated shared patterns and the estimated individual patterns pertaining to the data. In contrast to concatenated PCA, this method lacks estimation of what is truly joint and what is truly individual pertaining to the data sources in the analyses. Also, using separate individual PCAs for each data source may result in redundancies within each data source that could have been estimated as joint structure. Joint and Individual Variation Explained eliminates the weaknesses of both of these mentioned strategies and leads to accurate predictive results from multiple high-dimensional data sources.

In this article, we have considered JIVE as an extension of PCA for prediction from multiple sources of high-dimensional data. This general approach may also be extended to other data reduction techniques (eg, autoencoders[26] and non-negative matrix factorization[16]). Also, in this article, we have focused on the 2-step process of data reduction followed by predictive modeling. An advantage of such an approach is the flexibility to use a variety of classical statistical techniques (eg, logistic regression, Cox proportional hazards modeling) after the initial dimension reduction step. However, alternative single-step prediction approaches to assess the joint and individual predictive power of multisource high-dimensional data is an interesting direction of future work.

In this article, the JIVE ranks were estimated by permutation tests. Alternative approaches that are directly related to the predictive model, such as selecting the ranks via Bayesian information criterion or to give optimal predictive performance under cross-validation, may also be used. Alternative penalization strategies, such as an $L_1$ penalty term on the joint and individual singular values, are another direction of future work.

Jive.predict is a method for adding new patients into the currently JIVE-analyzed data sources an investigator or statistician may have at any given day. The swiftness of the algorithm is also a noticeable perk; the algorithm converges in less than 10 iterations. Instead of running a whole new JIVE analysis on the integrated data of the new and old subjects, one can bring in the raw data and use this new algorithm under 10 minutes. An additional advantage of jive.predict is that we do not need to construct an entirely new JIVE model for the data again, which is conceptually appealing. A well-validated prediction model does not need to change under new data. Once the algorithm quickly finishes, the investigator will obtain the scores and predict those sets of patients' outcomes, such as median survival time, probability of an event of cancer, and odds of being in a high-risk group of a disease.

### Acknowledgements

The authors would like to thank The Cancer Genome Atlas Research Network for providing data. They would also like to thank Michael O'Connell for his assistance with the R.JIVE package and Dr Xianghua Luo for fruitful discussions on survival modeling with high-dimensional covariates.

### Author Contributions

EFL conceived and designed JIVE and applications of JIVE for statistical modeling. AK analyzed the data and conceived and designed jive.predict and applications of jive.predict.

### REFERENCES

1. Massy WF. Principal components regression in exploratory statistical research. *J Am Stat Assoc*. 1965;60:234–256.
2. Bair E, Hastie T, Paul D, et al. Prediction by supervised principal components. *J Am Stat Assoc*. 2006;101:119–137.
3. Shen YJ, Huang SG. Improve survival prediction using principal components of gene expression data. *Genomics Proteomics Bioinformatics*. 2006;4:110–119.
4. Bild AH, Yao G, Chang JT, et al. Oncogenic pathway signatures in human cancers as a guide to targeted therapies. *Nature*. 2006;439:353–357.
5. Witten DM, Tibshirani R. Survival analysis with high-dimensional covariates. *Stat Methods Med Res*. 2010;19:29–51.
6. http://cancergenome.nih.gov/.
7. Brennan CW, Verhaak RG, McKenna A, et al. The somatic genomic landscape of glioblastoma. *Cell*. 2013;155:462–477.
8. Ohgaki H, Kleihues P. Epidemiology and etiology of gliomas. *Acta Neuropathol*. 2005;109:93–108.
9. Bleeker FE, Molenaar RJ, Leenstra S. Recent advances in the molecular understanding of glioblastoma. *J Neurooncol*. 2012;108:11–27.
10. Krex D, Klink B, Hartmann C, et al. Long-term survival with glioblastoma multiforme. *Brain*. 2007;130:2596–2606.
11. Zhao Q, Shi X, Xie Y, et al. Combining multidimensional genomic measurements for predicting cancer prognosis: observations from TCGA. *Brief Bioinform*. 2014;16:291–303.
12. Westerhuis JA, Kourti T, MacGregor JF. Analysis of multiblock and hierarchical PCA and PLS models. *J Chemometr*. 1998;12:301–321.
13. Ray P, Zheng L, Lucas J, et al. Bayesian joint analysis of heterogeneous genomics data. *Bioinformatics*. 2014;30:1370–1376.
14. Löfstedt T, Hoffman D, Trygg J. Global, local and unique decompositions in OnPLS for multiblock data analysis. *Anal Chim Acta*. 2013;791:13–24.
15. Schouteden M, Van Deun K, Wilderjans TF, et al. Performing DISCO-SCA to search for distinctive and common information in linked data. *Behav Res Methods*. 2014;46:576–587.
16. Yang Z, Michailidis G. A non-negative matrix factorization method for detecting modules in heterogeneous omics multi-modal data. *Bioinformatics*. 2016;32:1–8.
17. Lock EF, Hoadley KA, Marron JS, et al. Joint and individual variation explained (JIVE) for integrated analysis of multiple data types. *Ann Appl Stat*. 2013;7:523.
18. O'Connell MJ, Lock EF. R.JIVE for exploration of multi-source molecular data. *Bioinformatics*. 2016;32:2877–2879.
19. Wei Y. Integrative analyses of cancer data: a review from a statistical perspective. *Cancer Inform*. 2015;14:173–181.
20. Lock EF. *Vertical Integration of Multiple High-Dimensional Datasets* [PhD thesis]. Chapel Hill, NC: The University of North Carolina at Chapel Hill, 2012.
21. Cox D. Regression models and life tables. *J R Stat Soc*. 1972;34:187–220.
22. Leek JT, Johnson WE, Parker HS, et al. *SVA: Surrogate Variable Analysis*. R package version 3.18.0.
23. Benjamini Y, Hochberg Y. Controlling the false discovery rate: a practical and powerful approach to multiple testing. *J Roy Stat Soc B Met*. 1995;289–300.
24. R Core Team. *R: A Language and Environment for Statistical Computing*. Vienna, Austria: R Foundation for Statistical Computing; 2015. http://www.R-project.org/.
25. Akaike H. A new look at the statistical model identification. *IEEE T Automat Contr*. 1974;19:716–723.
26. Tan J, Ung M, Cheng C, et al. Unsupervised feature construction and knowledge extraction from genome-wide assays of breast cancer with denoising autoencoders. *Pac Symp Biocomput*. 2015; 20:132–143.

## Appendix 1

*Proof of score equality for jive.predict*

Let $X = X^*$ with given JIVE decompositions as follows:

$$X_1 = U_1 S + W_1 S_1 = U_1 S^* + W_1 S_1^* = X_1^*$$
$$X_2 = U_2 S + W_2 S_2 = U_2 S^* + W_2 S_2^* = X_2^*$$

Then, it follows that

$$S = S^*, S_1 = S_1^*, S_2 = S_2^*$$

where $rank(J_i + A_i) = rank(J_i) + rank(A_i)$, $J_i = U_i S$, and $A_i = W_i S_i$ due to $X$ following a JIVE decomposition.

*Proof*

As $rank(J_i + A_i) = rank(J_i) + rank(A_i)$

$$\Rightarrow col(J_i) \cap col(A_i) = \{0\} \quad (2)$$
$$(U_i S) \subseteq col(U_i) = col(J_i)$$
$$(W_i S_i) \subseteq col(W_i) = col(A_i)$$

$col(U_i)$ provides an orthonormal basis for $J_i$, and $col(W_i)$ provides an orthonormal basis for $A_i$. Take an arbitrary column of $X_1$ and $X_1^*$, say $\mathbf{x}_{\bullet j} \in X_1$ and $\mathbf{x}_{\bullet j}^* \in X_1^*$:

$$\mathbf{x}_{\bullet j} = U_1 \mathbf{s}_j + W_1 \mathbf{s}_{1j}$$
$$\mathbf{x}_{\bullet j}^* = U_1 \mathbf{s}_j^* + W_1 \mathbf{s}_{1j}^*$$
$$U_1 \mathbf{s}_j + W_1 \mathbf{s}_{1j} = U_1 \mathbf{s}_j^* + W_1 \mathbf{s}_{1j}^*$$
$$U_1 \mathbf{s}_j + W_1 \mathbf{s}_{1j} - (U_1 \mathbf{s}_j^* + W_1 \mathbf{s}_{1j}^*) = 0$$
$$U_1 (\mathbf{s}_j - \mathbf{s}_j^*) + W_1 (\mathbf{s}_{1j} - \mathbf{s}_{1j}^*) = 0$$
$$U_1 (\mathbf{s}_j - \mathbf{s}_j^*) \subseteq col(U_1) = col(J_1)$$
$$W_1 (\mathbf{s}_{1j} - \mathbf{s}_{1j}^*) \subseteq col(W_1) = col(A_1)$$

Therefore, these 2 products are linearly independent of each other due to equation (2):

$$U_1 (\mathbf{s}_j - \mathbf{s}_j^*) = 0$$
$$((U_1^T U_1)^{-1} U_1^T) U_1 (\mathbf{s}_j - \mathbf{s}_j^*) = ((U_1^T U_1)^{-1} U_1^T) 0$$
$$\mathbf{s}_j - \mathbf{s}_j^* = 0$$
$$\mathbf{s}_j = \mathbf{s}_j^*$$
$$W_1 (\mathbf{s}_{1j} - \mathbf{s}_{1j}^*) = 0$$
$$(W_1^T W_1)(\mathbf{s}_{1j} - \mathbf{s}_{1j}^*) = 0$$
$$\mathbf{s}_{1j} = \mathbf{s}_{1j}^*$$

An analogous argument can be used to show the equivalence of the scores for $X_2$ and $X_2^*$.